\documentclass[twocolumn,epjc3]{svjour3}          
\journalname{Eur. Phys. J. A}
\usepackage{cuted}
\usepackage{xcolor}
\usepackage{slashed}
\def\ly{\lambda_Y}
\def\nn{\nonumber\\}
\def\ba{\begin{eqnarray}}
\def\ea{\end{eqnarray}}
\usepackage{graphicx}
\usepackage{amsbsy}
\input{epsf}
\def\epa {\varepsilon  _\bot p_h}

\title{
Complete lowest order radiative corrections in semi-inclusive scattering of polarized particles
}
\author{ I. Akushevich\thanksref{addr1,e1}
            \and
H. Avakian\thanksref{addr2,e2}
            \and
     A. Ilyichev\thanksref{addr3,addr4,e3}
            \and
S. Srednyak\thanksref{addr1,e4}
}

\thankstext{e1}{e-mail: igor.akushevich@duke.edu}
\thankstext{e2}{e-mail: avakian@jlab.org}
\thankstext{e3}{e-mail: ily@hep.by}
\thankstext{e4}{e-mail: stan.sredn@gmail.com}

\institute{Physics Department, Duke University, Durham, NC 27708, USA\label{addr1}
\and
Thomas Jefferson National Accelerator Facility, Newport News, VA 23606, USA\label{addr2}
\and
 Belarusian State University,
220030  Minsk,  Belarus\label{addr3}
\and
Institute for Nuclear Problems,
 Belarusian State University,
220006  Minsk,  Belarus\label{addr4}
}

\date{Received: 30 June 2023 / Accepted: 04 October 2023}

\begin{document}

\maketitle
\begin{abstract}
The lowest order radiative corrections to the cross section and asymmetries measured in experiments on semi-inclusive deep inelastic scattering of polarized particles were calculated. Both exact and leading log expressions were presented and discussed for the total correction that include the contributions from the processes of (i) real photon emission with semi-inclusive processes, (ii) loop diagrams, and (iii) real photon emission with exclusive processes. Radiative corrections to the Sivers and Collins asymmetries in $\pi^+$ electroproduction were studied numerically within the kinematical conditions of modern experimental environments at Jefferson Laboratory (JLab). The Wandzura-Wilczek approximation for the semi-inclusive structure functions and MAID2007 parameterization for the six amplitudes of exclusive processes were used in numeric analyses. The results show that (i) radiative effects can generate a correction comparable to the size of Sivers and Collins asymmetries at the Born level, (ii) there is good agreement between the exact and leading-order corrections, (iii) external functions (that is, other than the Sivers and Collins functions in the respective asymmetries) can generate a contribution to the radiative correction up to 20\%, and (iv) there exists a strong dependence of the radiative correction on the models for semi-inclusive and exclusive structure functions.
\end{abstract}

\section{Introduction}
Modern achievements in theoretical physics as well as improvement of experimental techniques allow researchers to access the spin structure of the nucleon encoded in transverse momentum-dependent parton distribution functions by studying polarized semi-inclusive deep-inelastic scattering (SIDIS). The usual interpretation of the nucleon dynamics in high-energy interactions that is often limited to a simple one-dimensional picture of a fast moving nucleon has to be replaced by a truly 3-dimensional study of the nucleon structure \cite{Anselmino}.

The main task of data analysis in lepton nucleon scattering is to extract the basic contribution to the cross section that contains only one photon exchange process between lepton and hadron legs. However, this process is accompanied by other processes known as radiative corrections (RC) that cannot be distinguished from the basic process by experimental methods. These processes are related to the contributions of the additional virtual particles and real photon emission, and therefore, they are of the next order with respect to the QED fine structure constant $\alpha\approx 1/137$ and expected to be essentially suppressed due to its smallness. At Jefferson Lab three halls are involved in studies of SIDIS~\cite{Dudek:2012vr} including (i) the HMS and Super HMS at Hall C, (ii) the BigBite and Super BigBite, as well as, the SoLID detector at Hall A, and
(iii) CLAS12 at Hall-B with several experiments already approved to study in details the  modulations of the cross section in SIDIS, involving  azimuthal angles of hadrons ($\phi_h$) and nucleon spin 
($\phi_\eta$)\footnote{Note that due to specifics of our calculations we use the notation $\phi_\eta$ for the azimuthal angles of nucleon spin instead of generally accepted $\phi_S$} 
for
different hadron types, targets, and polarizations in a broad kinematic range. Measurements of all kind of  structure functions~\cite{Aram,Bacchetta} defined by corresponding azimuthal modulations containing $\cos\phi_h$ terms, like Sivers ($F_{UT}^{\sin(\phi_h-\phi_\eta)}$) and Collins ($F_{UT}^{\sin(\phi_h+\phi_\eta)}$), as well as Kotzinian-Multers ($F_{UL}^{\sin 2\phi_h}$),  Cahn ($F_{UU}^{\cos\phi_h}$), and Boer-Mulders ($F_{UU}^{\cos 2\phi_h}$) are expected to be most sensitive to RC due to large cosines generated by the Bethe-Heitler process due to shift of the direction of the virtual photon direction.
The newly achieved accuracies in the JLab experiments require renewed attention to RC calculations and their implementation in data analysis software. In this approach, RC have to be calculated within a theoretical model and extracted from the experimental data.

Usually with some exceptions (see, e.g. \cite{AI20221,AI20222}) RC to lepton-nucleon scattering is calculated in model-independent way \cite{ASh,MASCARAD}, when the calculation does not require additional assumptions on hadronic interactions. These RC include the radiation of an unobserved real photon from the lepton line, vacuum polarization, and lepton-photon vertex corrections. These effects can be calculated without any assumptions on hadron interactions and represent the so-called model-independent RC. They give the largest contributions to the total RC and can be calculated exactly or in the leading-log approximation if the accuracy provided by this approximation is sufficient. By ``exactly'' calculated RC we understand the analytic expressions obtained without any simplifying assumptions with opportunities for numeric estimates with any predetermined accuracy. The structure of dependence of the RC cross section on the electron mass is $\sigma_{RC}=A\log\frac{Q^2}{m^2}+B+O(m^2/Q^2)$, where $A$ and $B$ do not depend on the electron mass $m$. If only $A$ is kept in the formulae for RC, this is the leading log approximation. The leading log approximation can be sufficient in certain cases because the factor $\log{Q^2}/{m^2}$ is of the order of 15 for JLab energies.  The model independent RC include the effects of emissions of real and virtual photons from the lepton line only. Uncertainties of the model independent RC can come only from fits and data used for structure functions, whereas the model dependent corrections ({\it i. e.} box-type diagrams, emission by hadrons) require the additional information about hadron interactions and therefore contain additional purely theoretical uncertainties, which are hard to control.

Radiative corrections to the three-fold cross section over the Bjorken variables $x$, $y$ and the fraction of the virtual photon energy transferred to the detected hadron $z$ were estimated in \cite{SSh1,SSh2} and implemented in the SIRAD patch of the FORTRAN code POLRAD \cite{POLRAD}. The RC to the five-fold differential cross section for unpolarized particles with two additional variables characterized by the detected hadron (that is, the transverse momentum $p_t$ and the azimuthal angle between the lepton scattering and hadron production planes, $\phi_h$) was calculated in \cite{haprad}. These calculations did not contain the radiative tail from exclusive reactions as a separate contribution involving the exclusive
structure functions. The first estimate of the exclusive radiative tail contribution was made for the unpolarized SIDIS in \cite{AIO} and showed rather large effects in the region near the pion threshold. The explicit expressions for RC to SIDIS with initial polarized particles
were calculated in \cite{AI2019}. Based on the results obtained in this article, a Monte Carlo generator has been developed for the simulation of the hard-photon emission excluding the exclusive radiative tail \cite{SIDIS-RC}.   Recently RC to SIDIS in leading logarithmic approximation was estimated by Liu et al. in \cite{Liu} and our group in \cite{AIS2023}.

Here, we generalize the expressions obtained exactly and within the lowest order leading log approximation in \cite{AI2019} and \cite{AIS2023} respectively for polarized SIDIS. The first component of these developments is the inclusion of the contribution of the radiative tail from  exclusive processes or, simply, exclusive radiative tail. The contribution represents an important part of the total RC, but before the contribution was analyzed only for the case of unpolarized particles \cite{AIO}. Partly this is because of limited information on the exclusive structure functions for the spin-dependent part of the cross section. In this paper we represented the exclusive structure functions in the form appropriate for RC calculations in terms of the six invariant exclusive amplitudes \cite{maid0,maid1,maid}. The second is the analytic representation and numerical evaluation of the QED model independent RC. The obtained expressions could allow to obtain expressions for the RC with the highest accuracy achievable today: RC of the lowest order calculated exactly with the highest order RC obtained within the method of the electron structure functions \cite{Kuraev1,Kuraev2,ESFRAD1,ESFRAD2}. This approach requires the calculation of RC in the leading log approximation with further generalization using the methods of the electron structure functions. In turn, the leading log expressions for the cross section require the representation  of all kinematic variables in the kinematics of radiative processes (or in the so-called shifted kinematics). The explicit representation of the proton polarization appropriate for these expressions is the third component of our development.

The remainder of this article is organized as follows. The kinematics of SIDIS process, representation for the hadronic tensor, and the structure functions used in the literature, as well as the one-photon exchange (Born) contribution to the SIDIS process, are discussed in Sect.~\ref{kinborn}.  In Sect.~\ref{RCSIDIS} the results for RC are presented focusing on novel features in calculations of RC in SIDIS. The numerical evaluations of RC to Sivers \cite{Sivers} and Collins \cite{Collins} asymmetries using both exact calculation and leading logarithmic approximation are presented and discussed in Sect.~\ref{numres}. We presented the numeric illustrations for $\pi^+$-electroproduction at Jlab kinematic conditions, though the obtained results are rather general and can be applied to other hadron leptoproduction processes in polarized SIDIS. The Wandzura-Wilczek model \cite{WW} for SIDIS structure functions and the parameterization of MAID2007 for the six amplitudes of exclusive processes\cite{maid0,maid1,maid} were used in numeric analyzes.  Discussion and conclusion remarks are presented in the last Section \ref{conc}. Technical details are presented in three appendices.

\section{Kinematics and Born contribution}
\label{kinborn}
The  process of semi-inclusive hadron leptoproduction
\ba
l(k_1,\xi)+n(p,\eta)\longrightarrow l(k_2)+h(p_h)+x(p_x)
\ea
($k_1^2=k_2^2=m^2$, $p^2=M^2$, $p_h^2=m^2_h$) can be described by the following set of independent variables
\ba
&\displaystyle
x=-\frac{q^2}{2qp},\;
y=\frac{qp}{k_1p},\;
z=\frac{p_hp}{pq},\;
t=(q-p_h)^2,
\nn
&\displaystyle
\;
\phi_h,
\;
\phi.
\label{setvar}
\ea
Here $q=k_1-k_2$,
$\phi_h$ is the angle between
$({\bf k_1},{\bf k_2})$ and $({\bf q},{\bf p_h})$ planes,
and $\phi$ is the angle between
$({\bf k_1},{\bf k_2})$ and the ground
planes in the target rest frame reference system (${\bf p}=0$).

Also we use the following set of invariants:
\ba
&S=2pk_1,\;
Q^2=-q^2,\;
Q_m^2=Q^2+2m^2,\;
\nn&
X=2pk_2,\;
S_x=S-X,\;
S_p=S+X,
\nn&
\displaystyle
V_{1,2}=2k_{1,2}p_h,\;
V_+=\frac 1 2(V_1+ V_2),
\nn&
\displaystyle
V_-=\frac 1 2(V_1- V_2)=\frac 1 2(m_h^2-Q^2-t),
\nn&
S^\prime=2k_1(p+q-p_h)=S-Q^2-V_1,
\nn&
X^\prime=2k_2(p+q-p_h)=X+Q^2-V_2,
\nn&
W^2=(p+q)^2=S_x+M^2-Q^2,\;
\nn&
p_x^2=(p+q-p_h)^2=M^2+t+(1-z)S_x.
\nn&
\lambda_S=S^2-4M^2m^2,\;
\lambda_Y=S_x^2+4M^2Q^2,\;
\nn&
\lambda_1=Q^2(SX-M^2Q^2)-m^2\lambda_Y,\;
\nn&
\lambda_S^\prime=S^{\prime 2}-4m^2p_x^2,\;
\lambda_X^\prime=X^{\prime 2}-4m^2p_x^2.
\nn&
\lambda_m=Q^2(Q^2+4m^2).
\label{setvar2}
\ea

\begin{figure}[t]\centering
\scalebox{0.64}{\includegraphics{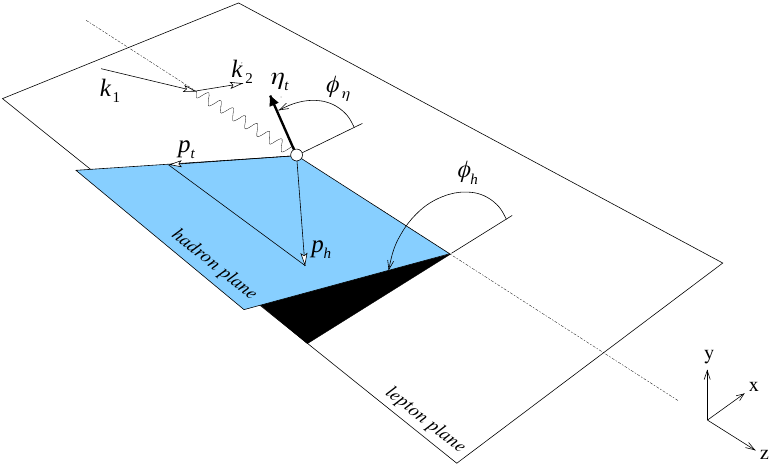}}
\caption{ Definition of azimuthal angles for semi-inclusive deep inelastic scattering in the target
rest frame \cite{Trento}. $p_t$ and $\eta _t$ are the transverse parts of $p_h$ and $\eta $ with respect to the photon
momentum.}
\label{fgtr}
\end{figure}

Following the Trento conventions \cite{Trento} we define the azimuthal angle $\phi_h$ of the outgoing hadron by
\ba
\cos \phi_h
=-\frac{k_1^\mu p_h^\nu g^t_{\mu\nu}}{k_t p_t}
=-\frac{k_2^\mu p_h^\nu g^t_{\mu\nu}}{k_t p_t},
\nn
\sin \phi_h
=-\frac{k_1^\mu p_h^\nu \varepsilon^t_{\mu\nu}}{k_t p_t}
=-\frac{k_2^\mu p_h^\nu \varepsilon^t_{\mu\nu}}{k_t p_t}.
\label{scph}
\ea
Here the tensors $\varepsilon^t_{\mu\nu}$ and $g^t_{\mu\nu}$
have nonzero components $\varepsilon^t_{12}=-\varepsilon^t_{21}$
and $g^t_{11}=g^t_{22}$ in the coordinate system of
Fig.~(\ref{fgtr}) and can be expressed through our variables as
\ba
\varepsilon^t_{\mu\nu}&=&\frac{2}{\sqrt{\ly}}\varepsilon_{\mu\nu\rho\sigma}p^\rho q^\sigma,
\nn
g^t_{\mu\nu}&=&\varepsilon^t_{\mu\rho}\varepsilon^\rho_{t\nu}=
g^\bot_{\mu\nu}-\frac{4Q^2}{\ly}p^\bot_\mu p^\bot_\nu,
\ea
where $g^\bot_{\mu\nu}=g_{\mu\nu}+q_\mu q_\nu/Q^2$ and for any four-vector $a_\mu^\bot=a_\mu+a q\; q_\mu /Q^2$.

The quantities
\ba
k_t&=&\sqrt{-g^t_{\mu \nu} k_1^\mu k_1^\nu}=\sqrt{-g^t_{\mu \nu} k_2^\mu k_2^\nu}=\sqrt{\frac{\lambda_1}{\lambda_Y}},\;
\nn
p_t&=&\sqrt{-g^t_{\mu \nu} p_{h\mu} p_{h\nu}}=\sqrt{p_{h0}^2-p_l^2-m_h^2}
\ea
are the transverse components of ${\bf k}_{1,2}$ and ${\bf p}_h$ with respect
to the virtual photon momentum, and  
\ba
p_{h0}=\frac{z S_x}{2M},
\qquad
p_l
=\frac{zS^2_x-4M^2V_-}{2M\sqrt{\lambda_Y}}
\ea
are the energy and the longitudinal component of the three-momenta of the detected hadron.

As a result,
\ba
\cos\phi_h&=&\frac{Q^2(zS_x S_p-4M^2V_+)-S_x(SV_2-XV_1)}{2p_t\sqrt{\ly\lambda_1}},
\nn
\sin\phi_h&=&-\frac{2\varepsilon ^{\mu\nu\rho\sigma}k_{1\mu}p_{h\nu}p_\rho q_\sigma}{p_t\sqrt{\lambda_1}}.
\ea

Since the initial lepton is considered to be longitudinally polarized, its
polarization vector has the form \cite{ASh,AI2019}
\begin{eqnarray}
\xi  =\frac {\lambda_e S}{m\sqrt{\lambda_S}}k_{1}-\frac{2\lambda_e m}{\sqrt{\lambda_S}}p_1.
\label{xi}
\end{eqnarray}

To describe the properties of initial hadron polarization, it is convenient to decompose the polarized vector of the proton on the complete orthogonal basis $({\bf x}_h,{\bf y}_h,{\bf z}_h)$. In this basis,  ${\bf z}_h$ is chosen along the direction of the virtual
photon three-momentum ${\bf q}={\bf k}_1-{\bf k}_2$, ${\bf x}_h$ lies in the plane $({\bf q},{\bf p}_h)$ along the part of the registered hadron momentum that is
transverse to the ${\bf z}_h$ axis whereas the rest axial axis is defined as ${\bf y}_h={\bf z}_h\times{\bf x}_h$.
 In covariant form its representation reads:
\ba
e^{h(0)}_\mu=\frac {p_\mu} M,
\qquad
\qquad
\qquad
&&e^{h(1)}_\mu=\frac 1{p_t}\Biggl[
p^\bot_{h \mu}
-
\frac {r_0}{\sqrt{\lambda_Y}}p^\bot_\mu
\Biggr],
\nonumber \\[1mm]
e^{h(2)}_\mu=-
2\frac{\varepsilon^{\mu \nu \rho \sigma}p_\nu q_\rho p_{h\sigma}}{p_t\sqrt{\lambda_Y}},
&&e^{h(3)}_\mu=\frac {2M^2 q_\mu-S_xp_\mu}{M\sqrt{\lambda_Y}},
\label{e1}
\ea
with  
\ba
r_0=\frac {2S_x(zQ^2+V_-)}{\sqrt{\lambda_Y}},
\label{r0}
\ea
and has the following properties
\ba
e^{h(a)}_\mu e^{h(b)}_\nu g^{\mu \nu}&=&g^{ab},
\nn
({\bf e}^{h(3)}\times {\bf e}^{h(1)})_\mu&=&\varepsilon_{\rho\mu\sigma\delta}
e^{h(0)}_\rho e^{h(3)}_\sigma e^{h(1)}_\delta=e^{h(2)}_\mu.
\label{e1pr}
\ea

As a result,
the target polarized vector $\eta$ can be decomposed on the basis presented in Eq.~(\ref{e1})
\ba
\eta=\sum_{i=1}^3\eta_i e^{h(i)}.
\label{etadec}
\ea
Here we take into account $\eta e^{h(0)}=M\eta p\equiv 0$.
Therefore, the components of the vector $\eta$ reads
\ba
\eta_1&=&-\eta e^{h(1)}=\eta_t \cos(\phi_\eta-\phi_h)=-\frac{p_h^\mu \eta^\nu g^t_{\mu \nu}}{p_t}
\nn
&=&-\frac{ p_h\eta+p_l \eta_3}{p_t},
\nn
\eta_2&=&-\eta e^{h(2)}=\eta_t \sin(\phi_\eta-\phi_h)=-\frac{p_h^\mu \eta^\nu \varepsilon^t_{\mu \nu}}{p_t}
\nn
&=&\frac{2\varepsilon_{\rho\sigma\gamma\delta}p^\rho p_h^\sigma q^\gamma \eta^ \delta}{p_t\sqrt{\ly}},
\nn
\eta_3&=&-\eta e^{h(3)}=-\frac{2M q\eta}{\sqrt{\ly}},
\nn
\eta_t&=&\sqrt{-g^t_{\mu\nu}\eta^\mu\eta^\nu}=\sqrt{1-\eta_3^2},
\label{etab}
\ea
where $\phi_\eta$ is an azimuthal angle defined in analogy to $\phi_h$ in Eqs.~(\ref{scph}), with $p_h$ replaced by $\eta$.

\begin{figure}[t]\centering
\vspace*{-6mm}
\scalebox{0.47}{\includegraphics{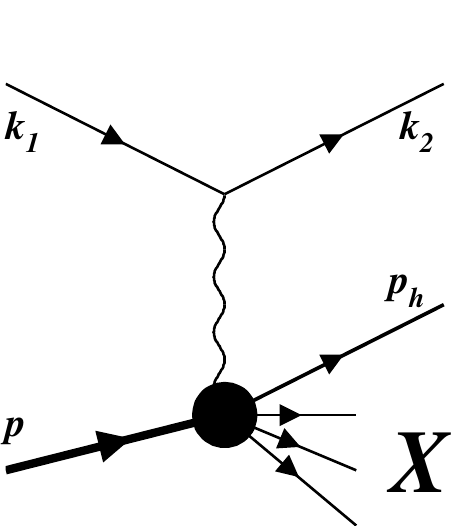}}
\caption{Feynman graph for the one-photon exchange (Born)
contribution to SIDIS scattering}
\label{fg1}
\end{figure}

The one-photon exchange (Born) contribution to SIDIS is presented by the Feynman graph in Fig.~\ref{fg1}
and has a form
\ba
d\sigma _B=\frac{(4\pi\alpha)^2}{2\sqrt{\lambda_S}Q^4}W_{\mu \nu}L^{\mu \nu}_B
d\Gamma _B,
\label{wl}
\ea
where the phase space is parametrized as
\ba
d\Gamma ^B&=&
(2\pi)^4\frac{d^3k_2}{(2\pi)^32k_{20}}
\frac{d^3p_h}{(2\pi)^32p_{h0}}
\nonumber\\
&=&\frac1{4(2\pi)^2}\frac {S S_xdx dy d\phi}{2\sqrt{\lambda_S}} \frac{S_xdzdp^2_td\phi_h}{4Mp_l}.
\ea

Taking into account (\ref{xi}) the leptonic tensor is
\begin{eqnarray}
L_B^{\mu \nu}&=&\frac 12 {\rm Tr}[({\slashed k}_2+m)\gamma_{\mu }({\slashed k}_1+m)(1+\gamma _5{\slashed \xi})\gamma_{\nu}]
\nonumber\\
&=&2[k_{1}^{\mu}k_{2}^{\nu}+k_{2}^{\mu}k_{1}^{\nu}-\frac {Q^2}2g^{\mu \nu}
\nonumber\\&&
+\frac {i\lambda_e}{\sqrt{\lambda_S}}
\varepsilon^{\mu \nu \rho \sigma}
(Sk_{2\rho} k_{1\sigma}+2m^2q_{\rho} p_{\sigma})
].
\label{lt0}
\end{eqnarray}

The covariant form for the hadronic tensor that was used in \cite{AI2019} for RC calculation is:
\ba
W_{\mu\nu}&=&\sum\limits_{i=1}^9w^i_{\mu\nu}{\cal H}_i
=-g^\bot_{\mu \nu} {\cal H}_1
+p^\bot_\mu p^\bot_\nu {\cal H}_2
\nonumber\\&&
+p^\bot_{h\mu} p^\bot_{h\nu} {\cal H}_3
+(p^\bot_{\mu} p^\bot_{h\nu}+p^\bot_{h\mu} p^\bot_{\nu}) {\cal H}_4
\nonumber\\&&
+i(p^\bot_{\mu} p^\bot_{h\nu}-p^\bot_{h\mu} p^\bot_{\nu}) {\cal H}_5
+(p^\bot_{\mu} n_{\nu}+n_{\mu} p^\bot_{\nu}) {\cal H}_6
\nonumber\\&&
+i(p^\bot_{\mu} n_{\nu}-n_{\mu} p^\bot_{\nu}) {\cal H}_7
+(p^\bot_{h\mu} n_{\nu}+n_{\mu} p^\bot_{h\nu}) {\cal H}_8
\nonumber\\&&
+i(p^\bot_{h\mu} n_{\nu}-n_{\mu} p^\bot_{h\nu}) {\cal H}_9,
\label{ht1}
\ea
where $n^{\mu}=\varepsilon^{\mu \nu \rho \sigma}q_\nu p_\rho p_{h\sigma}$
and the real nine  generalized structure functions ${\cal H}_i$
corresponding to a certain tensor structure $w^i_{\mu\nu}$. Each generalized
structure function ${\cal H}_i$ has two terms: ${\cal H}_{1-5}$ contain unpolarized components and terms proportional to $\eta_2$
components whereas the other structure functions include terms proportional to $\eta_1$ and  $\eta_3$. Therefore,  we obtain 5 spin-independent and 13 spin dependent real structure functions. In practice
 another set of this structure functions suggested in \cite{Bacchetta} is used:
\ba
{\cal H}_1&=&C_1[F_{UU,T}-F_{UU}^{\cos 2\phi_h}
+\eta_2(F_{UT}^{\sin (3\phi_h-\phi_\eta)}
\nn[2mm]&&
-F_{UT}^{\sin (\phi_h+\phi_\eta)}-F_{UT,T}^{\sin (\phi_h-\phi_\eta)})],
\nn[2mm]
{\cal H}_2&=&\frac{2C_1}{\ly p_t^2}\biggl[(r_0^2-2Q^2p_t^2)F_{UU}^{\cos 2\phi_h}
\nn[2mm]&&
+2p_tQ(r_0F_{UU}^{\cos \phi_h}
+p_tQ(F_{UU,T}+F_{UU,L}))
\nn[2mm]&&
+\eta_2((r_0^2-2Q^2p_t^2)(
F_{UT}^{\sin (\phi_h+\phi_\eta)}
-
F_{UT}^{\sin (3\phi_h-\phi_\eta)}
)
\nn[2mm]&&
+2p_tQ(r_0(
F_{UT}^{\sin \phi_\eta}
-
F_{UT}^{\sin (2\phi_h-\phi_\eta)}
)
\nn[2mm]&&
-p_tQ(
F_{UT,L}^{\sin (\phi_h-\phi_\eta)}+
F_{UT,T}^{\sin (\phi_h-\phi_\eta)}
)))\biggr],
\nn[2mm]
{\cal H}_3&=&\frac{2C_1}{p_t^2}\biggl[F_{UU}^{\cos 2\phi_h}
+\eta_2(
F_{UT}^{\sin (\phi_h+\phi_\eta)}
-
F_{UT}^{\sin (3\phi_h-\phi_\eta)}
)\biggr],
\nn[2mm]
{\cal H}_4&=&-\frac{2C_1}{\sqrt{\ly}p_t^2}\biggl[r_0F_{UU}^{\cos 2\phi_h}+p_tQF_{UU}^{\cos \phi_h}
\nn[2mm]&&
+\eta_2(p_tQ(F_{UT}^{\sin \phi_\eta}-F_{UT}^{\sin (2\phi_h-\phi_\eta)})
\nn[2mm]&&
+r_0(F_{UT}^{\sin (\phi_h+\phi_\eta)}-F_{UT}^{\sin (3\phi_h-\phi_\eta)}))\biggr],
\nn[2mm]
{\cal H}_5&=&-\frac{2C_1Q}{\sqrt{\ly} p_t}\biggl[F_{LU}^{\sin\phi_h}+\eta_2(F_{LT}^{\cos(2\phi_h-\phi_\eta)}-F_{LT}^{\cos\phi_\eta})\biggr],
\nn[2mm]
{\cal H}_6&=&\frac{2C_1}{\ly p_t^2}\biggl[\eta_1(
r_0(F_{UT}^{\sin (3\phi_h-\phi_\eta)}+F_{UT}^{\sin (\phi_h+\phi_\eta)})
\nn[2mm]&&
+2p_tQ(F_{UT}^{\sin \phi_\eta}+F_{UT}^{\sin (2\phi_h-\phi_\eta)}))
\nn[2mm]&&
+\eta_3(r_0F_{UL}^{\sin 2\phi_h}
+2p_tQF_{UL}^{\sin \phi_h})\biggr],
\nn[2mm]
{\cal H}_7&=&-\frac{2C_1}{\ly p_t^2}\biggl[\eta_1(2p_tQ(F_{LT}^{\cos(2\phi_h-\phi_\eta)}+F_{LT}^{\cos\phi_\eta})
\nn[2mm]&&
+r_0F_{LT}^{\cos(\phi_h-\phi_\eta)})
+\eta_3(r_0F_{LL}+2p_tQF_{LL}^{\cos\phi_h})\biggr]
\nn[2mm]
{\cal H}_8&=&-\frac{2C_1}{\sqrt{\ly }p_t^2}
\biggl[\eta_1(F_{UT}^{\sin (\phi_h+\phi_\eta)}+F_{UT}^{\sin (3\phi_h-\phi_\eta)})
\nn[2mm]&&
+\eta_3F_{UL}^{\sin 2\phi_h}\biggr],
\nn[2mm]
{\cal H}_9&=&\frac{2C_1}{\sqrt{\ly }p_t^2}
\biggl[\eta_1F_{LT}^{\cos(\phi_h-\phi_\eta)}
+\eta_3F_{LL}\biggr],
\ea
where $C_1=4Mp_l(Q^2+2xM^2)/Q^4$, and  $r_0$ is defined in Eq.~(\ref{r0}). The first and second subscripts in this set of the structure functions refer to the polarization of the initial lepton and proton,
respectively. The third index specifies the polarization of the virtual photon exchanged in the reaction. The upper indices show the sine/cosine to which the given structure function is proportional at the level of the Born cross section. The convenient parameterization in the Wandzura-Wilczek approximation was developed for this set of the structure functions \cite{WW}.

Finally, we find the Born contribution in the form of ref. \cite{AI2019}
\begin{eqnarray}
\sigma^B&\equiv &\frac{d\sigma^B}{dxdydzdp_t^2d\phi_hd\phi}
\nn
&=&
\frac{\alpha^2 SS^2_x}{8MQ^4p_l\lambda_S}\sum\limits_{i=1}^9\theta^B_i{\cal H}_i(\chi_i^1,\chi_i^2,Q^2,x,z,p_t),
\label{born}
\end{eqnarray}
where $\chi_{1-5}^1=1$, $\chi_{6-9}^1=\eta_1$, $\chi_{1-5}^2=\eta_2$, $\chi_{6-9}^2=\eta_3$, $\theta^B_i=L^{\mu \nu}w^i_{\mu \nu}/2$,
\ba
\theta^B_1&=&Q^2-2m^2,
\nn[1mm]
\theta^B_2&=&(SX-M^2Q^2)/2,
\nn[1mm]
\theta^B_3&=&(V_1V_2-m_h^2Q^2)/2,
\nn[1mm]
\theta^B_4&=&(SV_2+XV_1-zQ^2S_x)/2,
\nn[1mm]
\theta^B_5&=&\frac{2\lambda_eS\epa}{\sqrt{\lambda_S}} ,
\nn[1mm]
\theta^B_6&=&-S_p\epa ,
\nn[1mm]
\theta^B_7&=&
\frac {\lambda_eS}{4\sqrt{\lambda_S}}[\lambda_YV_+-S_pS_x(zQ^2+V_-)],
\nn[1mm]
\theta^B_8&=&-2V_+\epa ,
\nn[1mm]
\theta^B_9&=&\frac {\lambda_e}{2\sqrt{\lambda_S}}
[S (Q^2 (zS_x V_+-m_h^2S_p) + V_- (S V_2
-XV_1))
\nn[1mm]&&
+2m^2(4M^2V_-^2+\lambda_Y m_h^2
-z S_x^2 (z Q^2 +2 V_-))].
\nn
\label{thb}
\ea
Here
\ba
\varepsilon _\bot p_h=\varepsilon^{\mu \nu \rho \sigma}p_{h\;\nu} p_\nu  k_{1\rho }q_\sigma
=-\frac 12 p_t\sqrt{\lambda_1}\sin\phi_h.
\label{epst1}
\ea

\begin{figure}[t]\centering
\scalebox{0.47}{\includegraphics{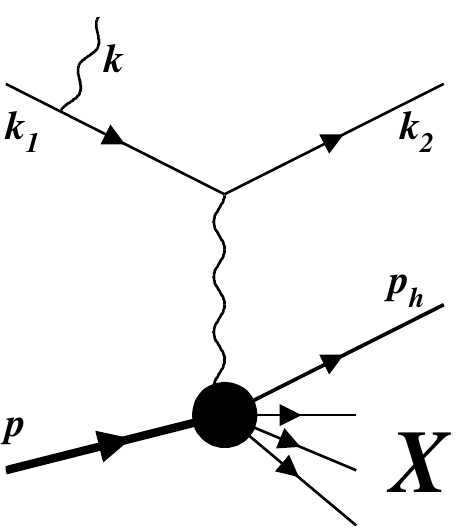}}
\scalebox{0.47}{\includegraphics{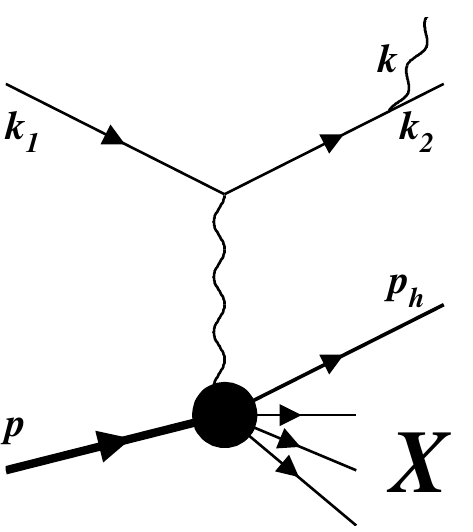}}
\\[-0.1cm]
{\bf \hspace{-.5cm} a) \hspace{3.3cm} b)\hspace{2.42cm}}
\\[0.1cm]
\scalebox{0.47}{\includegraphics{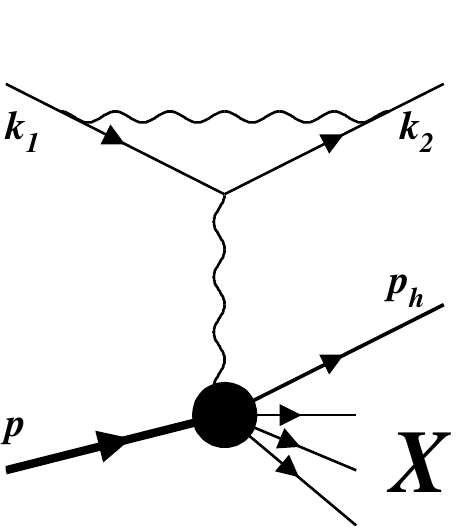}}
\scalebox{0.47}{\includegraphics{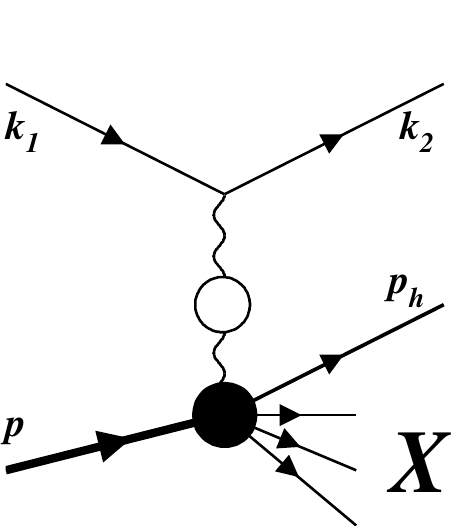}}
\\[-0.1cm]
{\bf \hspace{-.5cm} c) \hspace{3.3cm} d)\hspace{2.42cm}}
\\[0.1cm]
\scalebox{0.47}{\includegraphics{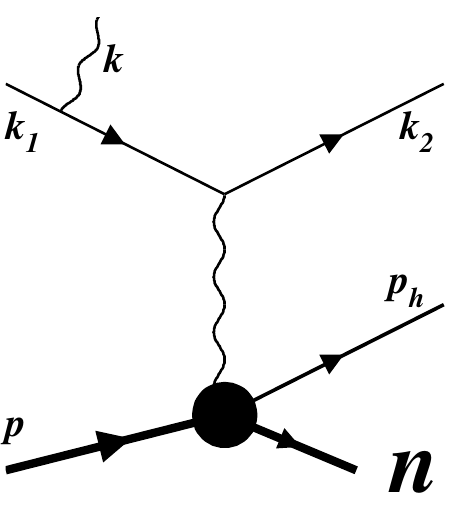}}
\scalebox{0.47}{\includegraphics{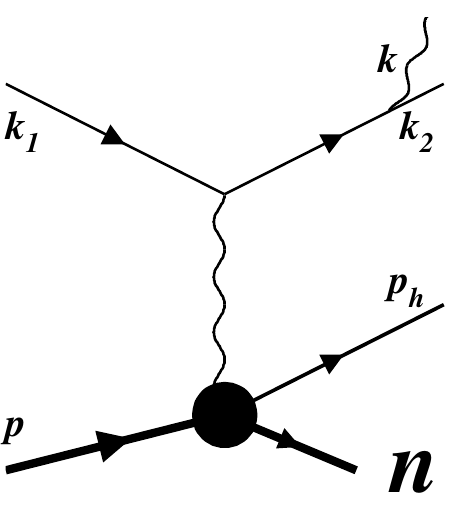}}
\\[-0.1cm]
{\bf \hspace{-.5cm} e) \hspace{3.3cm} f)\hspace{2.42cm}}
\\[0.1cm]
\caption{Feynman graphs for (a)-(d) SIDIS and (e), (f)  exclusive  radiative tail
contributions to the lowest order RC for SIDIS scattering}
\label{fgr}
\end{figure}

\section{Radiative corrections}
\label{RCSIDIS}
As it was presented in \cite{AI2019} RC to SIDIS consists of two parts, namely RC from the semi-inclusive contribution $\sigma^{in} $ presented by Feynman graphs in Fig.\ref{fgr}(a-d) and the
exclusive radiative tail $\sigma ^{ex}_r$  presented in Fig.\ref{fgr}(e,f).

Therefore, the radiatively corrected (or observed) contribution to the SIDIS cross section reads
\ba
\sigma^{obs}=
\sigma^B+
\sigma^{in}+
\sigma^{ex}_R.
\ea


\subsection{Semi-inclusive contribution}
According to \cite{AI2019}, the final expression for RC to SIDIS from semi-inclusive process reads
\ba
\sigma^{in}&=&
\frac{\alpha }{\pi }(\delta_{VR}
+\delta_{\rm vac}^l
+\delta_{\rm vac}^h
)\sigma^{B}
+\sigma^F_R
+\sigma^{\rm AMM},
\ea
 where the quantities representing the vacuum polarization by lepton and hadrons
($\delta_{\rm vac}^{l,h}$) and the contribution of anomalous magnetic moment $\sigma^{\rm AMM}$
 are described in \cite{AI2019}. The factor $\delta_{VR}$ is the result of cancellation of the infrared divergence, and $\sigma^F_R$ is the infrared-free contribution of the real photon radiation \cite{AI2019}. We present these in the form that is more compact (e.g., the new expression for the function $S_{\phi }$).  Thus,  
\ba
\delta_{VR}
&=&2(Q_m^2L_m-1)\log \frac{p_x^2-M_{th}^2}{m \sqrt{p_x^2}}
+\frac 12S^\prime L_{S^\prime}
+S_{\phi }
\nn &&
-2  
+\frac 12 X^\prime L_{X^\prime}
+\biggl( \frac 32 Q^2+4m^2 \biggr)L_m
\nn &&
-\frac{Q_m^2}{\sqrt{\lambda_m}}
\biggl(\frac 12\lambda_mL_m^2
+2{\rm Li}_2\biggl[\frac{2\sqrt{\lambda_m}}{Q^2+\sqrt{\lambda_m}} 
\biggr]
-\frac{\pi^2}2 
\biggr)
\nn
\label{dvr}
\ea
is the infrared free sum of the infrared divergent terms including soft photon emission and the vertex contribution.  
Here
\ba
L_m&=&\frac 1{\sqrt{\lambda_m}}\log\frac {\sqrt{\lambda_m}+Q^2}{\sqrt{\lambda_m}-Q^2},
\nonumber \\
L_{S^\prime}&=&\frac 1{\sqrt{\lambda_S^\prime}}\log\frac{S^\prime+\sqrt{\lambda_S^\prime}}{S^\prime-\sqrt{\lambda_S^\prime}},
\nonumber\\[1mm]
L_{X^\prime}&=&\frac 1{\sqrt{\lambda_X^\prime}}\log\frac{X^\prime+\sqrt{\lambda_X^\prime}}{X^\prime-\sqrt{\lambda_X^\prime}}
\label{lms}
\ea
and the explicit expression of the term $S_\phi$ is presented by Eq.~(40) of work \cite{AI2019} coincide with classical definition.
Recently a simpler expression of  $S_\phi$ was found \cite{lpcth}:  
\ba
&\displaystyle
S_\phi=
\frac{Q_m^2}{\sqrt{\lambda_m}}
\biggl (
\frac {\lambda_S^\prime}4 L^2_{S^\prime}-
\frac {\lambda_X^\prime}4 L^2_{X^\prime}
\nn[2mm]
&\displaystyle
+
{\rm Li}_2\biggl[1-
\frac {\rho}{(S^\prime+\sqrt{\lambda_S^\prime})}\biggr]
+{\rm Li}_2\biggl[1-
\frac {(S^\prime+\sqrt{\lambda_S^\prime})\rho}{4m^2p_x^2}\biggr]
\nn[2mm]
&\displaystyle
-
{\rm Li}_2\biggl[1-
\frac {Q^2(X^\prime+\sqrt{\lambda_X^\prime})\rho}{p_x^2(Q^2+\sqrt{\lambda_m})^2}\biggr]
\nonumber \\
&\displaystyle
-
{\rm Li}_2\biggl[1-
\frac {4m^2Q^2\rho}{(X^\prime+\sqrt{\lambda_X^\prime})(Q^2+\sqrt{\lambda_m})^2}\biggr]
\biggr),
\label{sph0}
\ea
where
\begin{eqnarray}
{\rm Li}_2(x)=-\int\limits^x_0\frac{\log|1-y|}y dy
\end{eqnarray}
is Spence's dilogarithm and
\begin{eqnarray}
\rho=\frac{(Q^2_m+\sqrt{\lambda_m})S^\prime-2m^2X^\prime }{\sqrt{\lambda_m}}.
\end{eqnarray}

The finite part of real photon emission
\ba
e(k_1,\xi)+n(p, \eta)\to e(k_2)+h(p_h)+x(p_x)+\gamma(k),
\ea
has a form
\ba
\sigma_R^F&=&-\frac{\alpha ^3S S_x^2}{64\pi ^2M p_l\lambda_S\sqrt{\lambda _Y}
}
\int\limits_{\tau _{\rm min}}^{\tau _{\rm max}}
d\tau  
\int\limits_{0}^{2\pi}
d\phi_k  
\int\limits_{0}^{R _{\rm max}}
dR  
\nn[1mm]&&\times
\sum\limits_{i=1}^9
\Biggl[
\sum\limits_{j=1}^{k_i}\frac{{\cal H}_i(\tilde \chi_i^1,\tilde \chi_i^2,Q^2+\tau R, \tilde x, \tilde z, \tilde p_t)\theta_{ij}R^{j-2}}{(Q^2+\tau R)^2}
\nn[1mm]&&\qquad
-
\frac{\theta_{i1}}R
\frac {{\cal H}_i(\chi_i^1,\chi_i^2,Q^2,x,z,p_t)}{Q^4}
\Biggr],
\label{srfin}
\ea
where $\tilde \chi_{1-5}^1=1$, $\tilde \chi_{6-9}^1=\tilde \eta_1$, $\tilde \chi_{1-5}^2=\tilde \eta_2$, $\tilde \chi_{6-9}^2=\tilde \eta_3$, $k_i=\{3,3,3,3,3,4,4,4,4\}$,  
\ba
R=2kp,\qquad \tau=kq/kp
\ea
and $\phi_k$ is an angle between $({\bf k}_1,{\bf k}_2)$ and $({\bf k},{\bf q})$ planes.
The limits of integration are
\ba
R_{\rm max}=\frac {p_x^2-M_{th}^2}{1+\tau-\mu },
\qquad
\tau _{\rm max/min}=\frac{S_x\pm \sqrt{\lambda_Y}}{2M^2},
\ea
and the variable $\mu $ is defined as
\ba
\mu&=&\frac{kp_h}{kp}=\frac{p_{h0}}{M}+\frac{p_l(2\tau M^2-S_x)}{M\sqrt{\lambda_Y}}
\nn&&
-2Mp_t\cos(\phi_h-\phi_k)\sqrt{\frac{(\tau_{\rm max}-\tau)(\tau-\tau_{\rm min})}{\lambda_Y}}.
\label{mudef}
\end{eqnarray}

The variables with the tilde symbol are defined as
\ba
&\displaystyle
\tilde x=\frac {Q^2+\tau R}{S_x-R},\;
\tilde z=\frac {zS_x}{S_x-R},\;
\nn
&\displaystyle
\tilde p_t=\sqrt{-\tilde g^t_{\mu\nu}p_h^\mu p_h^\nu}=\sqrt{p_{h0}^2-\tilde p_l^2-m_h^2},
\nn
&\displaystyle
\tilde p_l=\frac{z S_x(S_x-R)-2M^2(2V_--\mu R)}{2M\sqrt{\tilde \ly}},
\nn
&\displaystyle
\tilde \ly=(S_x-R)^2+4M^2(Q^2+\tau R),
\ea
and $\tilde g^t_{\mu\nu}$ as well as the components of the target polarized vector $\tilde \eta_i$ are presented      in \ref{etat}.

The explicit expression for the quantities $\theta_{ij}$ can be found in Appendix B of ref.~\cite{AI2019}. In \ref{th5} the improved expression for one of them is presented and discussed.

\subsection{Exclusive radiative tail}
Similarly to  SIDIS process, the exclusive hadron leptoproduction $\gamma^*  + n \to h + n^\prime$ with the initial polarized nucleon can be described by
5 spin-independent and 13 spin-dependent structure functions. For their representation we use the hadronic tensor in the form (\ref{ht1}) with 9 generalized exclusive structure functions that depend on three variables $W^2$, $Q^2$ and $t$, namely ${\mathcal H}^{ex}_i(\chi_i^{1ex},\chi_i^{2ex},W^2,Q^2,t)$.
Here, similarly to the semi-inclusive process, $\chi_{1-5}^{1ex}=1$, $\chi_{6-9}^{1ex}=\eta^{ex}_1$, $\chi_{1-5}^{2ex}=\eta^{ex}_2$, $\chi_{6-9}^{2ex}=\eta^{ex}_3$.

Using these definitions   the exclusive radiative tail
can be presented as double integral over $\phi_k$ and $\tau$ \cite{AI2019}:
\ba
&\displaystyle
\sigma^{ex}_{R}=-\frac{\alpha^3SS_x^2}{2^9\pi^5Mp_l\lambda_S\sqrt{\lambda_Y}}
\int\limits_{\tau _{\rm min}}^{\tau _{\rm max}}d\tau
\int\limits_0^{2\pi}d\phi_k
\nonumber \\
&\displaystyle
\times
\sum_{i=1}^{9}
\sum_{j=1}^{k_i}
\frac{{\mathcal H}^{ex}_i(\tilde\chi_i^{1ex},\tilde\chi_i^{2ex}, Q^2+R_{er}\tau,{\tilde W}^2, \tilde t)\theta_{ij}R_{ex}^{j-2}}{(1+\tau-\mu)\tilde Q^4},
\nn
\label{sre1}
\ea
where $m_u$ is the mass of undetected hadron,
\ba
\tilde W^2=W^2-(1+\tau )R_{ex},
\qquad
\tilde t=t+(\mu-\tau)R_{ex}.
\ea
and the third photonic variable is fixed by the semi-inclusive variable $z$
\ba
&\displaystyle
R_{ex}=\frac{p_x^2-m_u^2}{1+\tau-\mu}.
\ea
Therefore, the components of the target polarized vector for the exclusive radiative tail $\tilde \eta^{ex}_{1-3}$ are obtained from the respective SIDIS quantities presented in \ref{etat} with the replacements $R\to R_{ex}$.

In the present paper, we are interested in the electroproduction of pions
that is described by the six invariant amplitudes $A_{1-6}$ introduced in \cite{maid0}. Their numerical values in the resonance region (i.e., for $0<W^2<4$ GeV$^2$ and $0<Q^2<5$  GeV$^2$)
are obtained from MAID2007~\cite{maid}, and the asymptotic extension in the regions with the highest values of these quantities is obtained using the fit from \cite{Browman:1975fr}.

On the level of one photon exchange (Born) the matrix elements for the exclusive leptoproduction can be expressed in terms of these amplitudes \cite{maid1} as:
\ba
{\cal M}^\mu
=\bar U(p_u)
\Gamma_{ex}^\mu U(p)
=\bar U(p_u)
\left(\sum\limits_{i=1}^6 \Gamma_i^\mu A_i\right)U(p),
\label{mexcl}
\ea
where
\ba
\Gamma_1^\mu&=&\frac{i}{2}\gamma_5\left(\slashed{q}\gamma^\mu-\gamma^\mu \slashed{q}\right),
\nn
\Gamma_2^\mu&=&i\gamma_5\left[\left(2 V_-+Q^2\right)p^\mu
- S_x\left(p_h^\mu-\frac{1}{2}q^\mu\right)\right],
\nn
\Gamma_3^\mu&=&i\gamma_5\left(\slashed{q}p_h^\mu-V_-\gamma^\mu\right),
\nn
\Gamma_4^\mu&=&i\gamma_5\left(2\slashed{q}p^\mu-S_x\gamma^\mu \right)-2M \Gamma_1^\mu,
\nn
\Gamma_5^\mu&=&i\gamma_5\left(q^\mu V_-+p_h^\mu Q^2 \right),
\nn
\Gamma_6^\mu&=&-i\gamma_5\left(\slashed{q} q^\mu+\gamma^\mu Q^2\right).
\label{mexcl1}
\ea

As a result, the exclusive hadronic tensor reads:
\ba
W^{\mu\nu}_{ex}=-\frac {{\rm Tr}
\left[
\Gamma_{ex}^\mu (\slashed{p}+M)(1+\gamma_5\slashed{\eta})
 \bar \Gamma_{ex}^\nu (\slashed{p}_u+M)
\right]}{8\pi \alpha},
\label{wex}
\ea
where $\bar \Gamma_{ex}^\nu=\gamma_0\Gamma_{ex}^{\nu \dag} \gamma_0$, and $\Gamma_{ex}^\nu$ is defined by Eq.~(\ref{mexcl1}).

The result of calculation of the traces in (\ref{wex}) can be presented in the form of the standard hadronic tensor (\ref{ht1}) with the coefficients at the tensor structures representing the  
 generalized exclusive structure functions of exclusive processes that contribute to the cross section of exclusive radiative tail (\ref{sre1}):
\ba
{\cal H}^{ex}_1&=&H_{22}^{ex},
\nn
{\cal H}^{ex}_2&=&\frac 1{\ly}\biggl[4Q^2(H_{00}^{ex}+H_{22}^{ex}) -\frac{4 Q r_{ex}}{p_t}H_{01}^{ex\;r}
\nn&&
+r_{ex}^2\frac{H_{11}^{ex}-H_{22}^{ex}}{p_t^2}\biggr],
\nn
{\cal H}^{ex}_3&=&\frac{H_{11}^{ex}-H_{22}^{ex}}{p_t^2},
\nn
{\cal H}^{ex}_4&=&\frac 1{\sqrt{\ly}}\biggl[r_{ex}\frac{H_{22}^{ex}-H_{11}^{ex}}{p_t^2}+\frac{2Q}{p_t}H_{01}^{ex\;r}\biggl],
\nn
{\cal H}^{ex}_5&=&\frac{2Q}{p_t\sqrt{\ly}}H_{01}^{ex\;i},
\nn
{\cal H}^{ex}_6&=&\frac{2}{p_t\ly}\biggl[2Q H_{02}^{ex\;r}-\frac{r_{ex}}{p_t}H_{12}^{ex\;r}\biggr],
\nn
{\cal H}^{ex}_7&=&\frac{2}{p_t\ly}\biggl[2Q H_{02}^{ex\;i}-\frac{r_{ex}}{p_t}H_{12}^{ex\;i}\biggr],
\nn
{\cal H}^{ex}_8&=&\frac{2}{p_t^2\sqrt{\ly}}H_{12}^{ex\;r},
\nn
{\cal H}^{ex}_9&=&\frac{2}{p_t^2\sqrt{\ly}}H_{12}^{ex\;r},
\label{hi}
\ea
where $Q=\sqrt{Q^2}$ and
\ba
r_{ex}=\frac{2(Q^2(M^2-m_u^2+S_x+t)+S_xV_{-})}{\sqrt{\ly}}.
\ea
These structure functions ${\cal H}^{ex}_i$ depends on amplitudes $A_i$ and the components of the polarization vector $\eta$ but these dependencies are hidden in $H^{ex\;r,i}_{ab}$ and explicitly presented in \ref{exsf}. These expressions show that we have 5 unpolarized and 13 spin dependent structure functions exactly as in the case of SIDIS.

\section{Numerical results}
\label{numres}

Numerical analysis is focused on the evaluation of RC to the Sivers $A_{UT}^{\sin(\phi_h-\phi_\eta )}$ and Collins $A_{UT}^{\sin(\phi_h+\phi_\eta )}$ asymmetries for electroproduction of $\pi^+$ in the kinematics of JLab experiments. These asymmetries are defined through the ratio of two-fold integrals:
\ba
A_{UT}^{\sin(\phi_h\pm\phi_\eta)}=
\frac{2\displaystyle\int\limits_0^{2\pi}d\phi_h\int\limits_0^{2\pi}d\phi_\eta \sin(\phi_h\pm\phi_\eta)\sigma}
{\displaystyle\int\limits_0^{2\pi}d\phi_h\int\limits_0^{2\pi}d\phi_\eta \sigma}.
\label{scasym}
\ea

Recall that the exclusive radiative tail (\ref{sre1}) will be calculated using the six complex amplitudes whose numerical values are obtained from the MAID2007 website \cite{maid}.

For semi-inclusive structure functions, the Wand\-zura-Wilczek type approximation \cite{WW} is used, which is based on the set of structure functions that was introduced in \cite{Bacchetta} and discussed in Sect.~\ref{kinborn}. Only 15 of them are nonzero ($F_{UU,L}=0$, $F_{LU}^{\sin\phi_h}=0$ and $F_{UT,L}^{\sin(\phi_h-\phi_\eta)}=0$).

At the one photon exchange (Born) level, the  Sivers and Collins asymmetries in Wandzura-Wilczek model (\ref{scasym}) are expressed as the  ratios of $F_{UT,T}^{\sin(\phi_h-\phi_\eta)}$ and  $F_{UT}^{\sin(\phi_h+\phi_\eta)}$ to $F_{UU,T}$:
\ba
A_{UT}^{\sin(\phi_h-\phi_\eta)}&=&\frac{F_{UT,T}^{\sin(\phi_h-\phi_\eta)}}{F_{UU,T}},
\nn
A_{UT}^{\sin(\phi_h+\phi_\eta)}&=&\frac{1-y}{1-y+y^2/2}\frac{F_{UT}^{\sin(\phi_h+\phi_\eta)}}{F_{UU,T}}.
\label{scasym2}
\ea
Here $F_{UT,T}^{\sin(\phi_h-\phi_\eta)}$ is Sivers \cite{Sivers} structure functions which describes
the distribution of unpolarized quarks inside a transversely polarized proton, $F_{UT}^{\sin(\phi_h+\phi_\eta)}$ is Collins \cite{Collins} fragmentation function, which decodes
the fundamental correlation between the transverse spin of a fragmenting quark and the transverse momentum of the produced final hadron,  and $F_{UU,T}$ is the structure function due to transverse polarization of the virtual photon.

\begin{figure*}[t]
\hspace*{5mm}
\scalebox{0.75}{\includegraphics{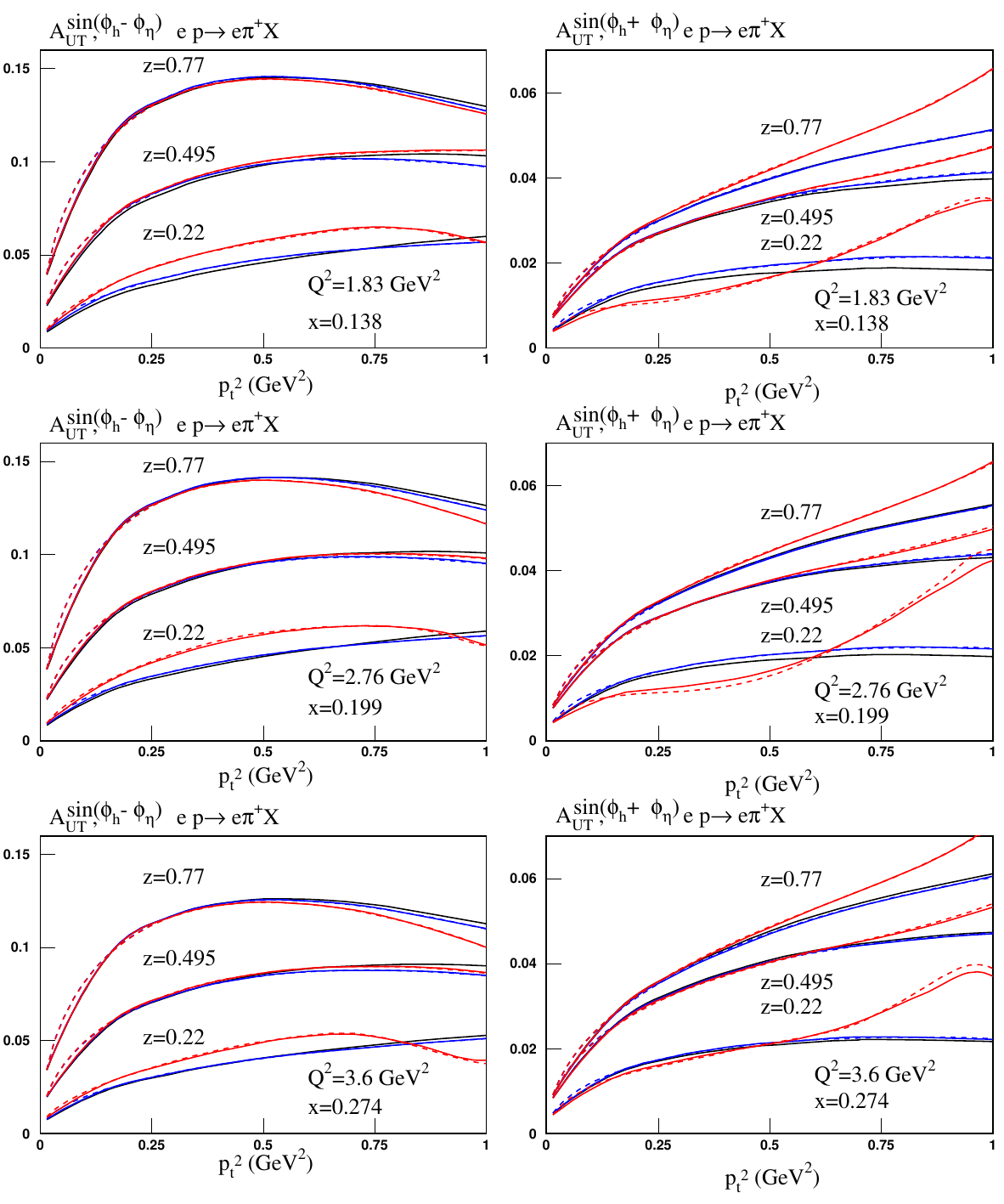}}
\caption{$p_t$-dependence of Sivers (left) and Collins (right) asymmetries for $\pi^+$ electroproduction. Black, blue, and red lines correspond to Born, Born neglecting the exclusive contributions,  and total contributions to these asymmetries, respectively. Solid (dashed) lines correspond to exact RC (RC in the leading logarithmic approximation).    
}
\label{fig4}
\end{figure*}

We evaluated RC for Sivers and Collins asymmetries both exactly and in the lowest order leading log approximation. $p_t^2$-dependence of these asymmetries is presented in Fig.~\ref{fig4} under the JLab kinematic conditions with the electron beam energy $E_{beam}=10.65$ GeV. The black, blue, and red curves in Fig.~\ref{fig4} correspond to the Born, Born with RC excluding the contribution of the exclusive radiative tail,  and total contributions, respectively. Solid and dashed lines correspond to exact RC and RC in the leading logarithmic approximation. The results show that the total radiative corrections have a rather complicated $p_t$ dependence due to the contribution of the exclusive radiative tail. This is especially pronounced for the region of small $z$. The latter is expected because the kinematic room for the radiated photon is higher for small $z$ (\ref{setvar2}). Collins asymmetry is more sensitive to this effect compared to Sivers asymmetry. We note that the contribution of the exclusive radiative tail to the Collins asymmetry can exceed the asymmetry calculated in the Born approximation. The region of larger $z$ and large $p_t$ is close to the pion threshold. In this region $p_x^2$ is close to $M_{th}^2$ and therefore the first term of (\ref{dvr}) can diverge. This divergence cancels by adding the effect of multiple soft photon emissions through the exponentiation procedure suggested by Yennie, Frautschi and Suura  \cite{YFS} and further developed and applied for this types of divergence by Shumeiko \cite{Sh}.
 Our general conclusion form these studies is that the results show rather good agreement between the exact and leading-order corrections in the kinematics of JLab experiments.

We also tested the magnitude of the effects on the Sivers and Collins asymmetries generated by other structure functions when we exclude the contribution from the leading SIDIS structure functions of the respective asymmetries, i. e., $F_{UT,T}^{\sin(\phi_h-\phi_\eta)}$ for Sivers and $F_{UT}^{\sin(\phi_h+\phi_\eta)}$ for Collins asymmetries. At the Born level these asymmetries (\ref{scasym2}) are exactly zero, so it is interesting to evaluate the effect generated by other structure functions due to RC.  Numerical estimation shows that these effects can reach 3\% for the Sivers asymmetries and 10\% for the Collins asymmetries. These effects
increase with decreasing $Q^2$. We also detected a strong model dependence on these effects. For example, if we modify the structure function $F_{UT}^{\sin\phi_\eta}=F_{UT,T}^{\sin(\phi_h-\phi_\eta)}$ to make it more appropriate for HERMES measurements \cite{HERMES}, the effect will be greater: about 20\% for the Sivers and 60\% Collins asymmetries.


\section{Discussion and Conclusion}
\label{conc}
The complete RC of the lowest order includes the contributions of the loops and real photon emission. The effects of loops are combined to the cross section of soft photon emission resulting in the infrared divergence free correction that is factorized at the Born cross section. Radiation of the hard photon results in two contributions depending on the type of hadronic process: semi-inclusive or pure exclusive channel. The second process provides a contribution to the Born SIDIS cross section because the unobserved final state (the hard photon and the exclusive hadron) can be kinematically similar to the final unobserved hadron state in the SIDIS process. Both SIDIS and exclusive processes are described by 5 spin-independent and 13 spin-dependent structure functions, and limited knowledge on these functions (especially in the case of scattering of polarized particles) represent the main obstacle in implementation of RC to SIDIS cross sections and spin asymmetries in actual experimental data analysis. In this paper we firstly demonstrated how the complete RC can be calculated in kinematical points of the modern experiments at JLab. We presented the minimal but complete set of analytic expressions and developed an implementation scheme for these 18 structure functions. For the case of exclusive scattering this was done for the first time.

All contributions to the RC including the exclusive radiative tail were calculated both exactly and in the lowest order leading-log approximation that generalize previous calculations in \cite{AIS2023}. The analytic formulae are presented in a convenient form when the polarization vector of the proton is explicitly expressed in terms of the kinematic variables of the base SIDIS process and integration variables. The explicit representation of the proton polarization vector in the kinematics of radiative processes is one important component of this development.  These results are presented in \ref{etat} for the shifted kinematic both for exact RC and RC in leading log approximation. For example, we demonstrated that in the leading log approximation the shifted components of the polarization vector can be expressed as a composition of three simple rotations.  Another innovation in our analysis is  the specific implementations of the exclusive structure functions originally provided by the six complex invariant amplitudes $A_{1-6}$ introduced in \cite{maid0}. The numerical value for these amplitudes in the resonance region (i.e., for $0<W^2<4$ GeV$^2$ and $0<Q^2<5$  GeV$^2$) are obtained from MAID2007~\cite{maid}, and the asymptotic extension in the regions with the highest values of these quantities is obtained using the fit from \cite{Browman:1975fr}. The matrix elements for the exclusive leptoproduction (\ref{mexcl},\ref{mexcl1}) allowed us to construct the exclusive hadronic tensor in the form (\ref{wex}) and extract necessary structure functions.  However for the representation of the exclusive structure functions  in \ref{exsf} we use Chew-Goldberger-Low-Nambu amplitudes \cite{CGLN} ${\cal F}_i$. The Wandzura-Wilczek model \cite{WW} for SIDIS structure functions is used, and the set of these  for SIDIS structure functions and MAID-based exclusive structure functions \cite{maid0,maid1,maid} represents the minimal set of structure functions for data analyses in experiments on SIDIS measurements.
 
The obtained results are rather general and can be applied to any hadron leptoproduction in polarized SIDIS.  However, as a numerical illustration
we restricted our consideration only to $\pi^+$-electroproduction at Jlab kinematic conditions when the unpolarized electron is scattering off the transversely polarized proton. The numerical analysis was focused on the Sivers \cite{Sivers} and Collins \cite{Collins} asymmetries, as both asymmetries are key observables in studies of the 3D structure of the nucleon at JLab and future Electron Ion Collider.
Comparison of the exact and leading-order corrections shows good agreement.  RC increases with increasing $p_t$ of the detected hadron. The exclusive radiative tail gives a large contribution for small $z$ and this effect requires further investigation, including quantifying the contributions of possible resonances and model dependence. Similarly to the unpolarized case, the exclusive radiative tail gives the maximum contribution in the region of high $p_t$, that is, when $p_t$ goes to the maximum value allowed by kinematics of the SIDIS process. This region is close to the pion threshold, so further steps in the analysis of RC in this region should include the exponentiation procedure to account for multiple-photon emission \cite{YFS,Sh}. The effect of non-leading structure functions was found to be quite moderate (which does not exceed 3\% for the Sivers and 10\% for Collins asymmetries). These estimates are sensitive to the model for structure functions. Studies of model dependence should be an important step in further improving RC calculations in SIDIS experiments.

Both exact and leading log formulae were presented in this paper. There are several reasons why we need to consider the approximation. Experimentalists continue using approximate formulae in their analyses and sometimes these approximations are not so justified, e.g., the soft photon approximation. That is why we usually present the set of exact formulae and reasonably approximated expressions in the leading log approximation. However there are at least two theoretical reasons why we need to use leading log approximation in our paper. First, these formulae represent an intermediate step for the application of the methods of the electron structure functions and the task ``to represent formulae for QED RC with maximal accuracy available for now''. Currently, this task is resolved by representing the lowest order RC exactly (in the meaning we discussed above) and the effects of higher order RC using the methods of the electron structure functions.  Second, the leading log formulae provide certain factorization, i.e., the RC cross section is expressed in terms of the Born cross section, and this essentially extends its area of applicability in both experimental and pure theoretical studies. For example, the form is identical for unpolarized and polarized parts of RC. Experimentalists appreciate this form because it helps them to understand why RC in certain regions is too large or small. This is especially important for SIDIS measurements where the cross section is five-fold and each kinematical variable produces their unique and often rapidly changing contribution to the Born cross section.

The points that have to be addressed in further studies also include: i) implementation of other models for structure functions for both semi-inclusive and exclusive processes, ii) implementation of RC procedure of experimental data that is important to reduce the bias in extracted asymmetries; this bias is proportional to the difference between the values of asymmetries (finally extracted and used in RC codes) in a given bin, iii) finalize the analytic and numeric comparison of leptonic RC calculated by other groups, e.g., the calculation of Liu et al. \cite{Liu}, iv) discuss and decide whether and how non leptonic corrections (including box diagrams and emission by hadrons) should be calculated and implemented, v) implement approaches for approximate or even exact calculations of the higher order corrections with paying a specific attention to the region close to the pion threshold, vi) {estimation of high order effects using electron structure function approach \cite{Kuraev1,Kuraev2,ESFRAD1,ESFRAD2}, and vii) continue developing Monte Carlo generators, e.g., implement the exclusive radiative tail to currently available generators, e.g., \cite{SIDIS-RC}.

\section*{Acknowledgements}
This material is based upon work supported by the U.S. Department of Energy, Office of Science, Office of Nuclear Physics under contract DE-AC05-06OR23177.

\appendix
\section{Components of $\tilde \eta _i$}
\label{etat}
The components of the target polarized vector $\tilde \eta_i$ that contributes to Eq.~(\ref{srfin}) have to be defined in respect of vector $q-k$ instead of $q$ as in Eq.~(\ref{etab}) in a following way:
\ba
\tilde \eta_1&=&\tilde \eta_t \cos(\tilde \phi_\eta-\tilde \phi_h)=-\frac{p_h^\mu \eta^\nu \tilde g^t_{\mu \nu}}{\tilde p_t}
\nn
&=&-\frac{p_h\eta+\tilde p_l \tilde \eta_3}{\tilde p_t}
\nn
&=&\frac 1{\tilde p_t}\biggl[\eta_3 p_l+\eta_1 p_t-\frac{\tilde p_l(2M k\eta+\eta_3\sqrt{\ly})}{\sqrt{\tilde \ly}}\biggr],
\nn
\tilde \eta_2&=&\tilde \eta_t \sin(\tilde \phi_\eta-\tilde \phi_h)=-\frac{p_h^\mu \eta^\nu \tilde \varepsilon^t_{\mu \nu}}{\tilde  p_t}
\nn
&=&\frac{2\varepsilon_{\rho\sigma\gamma\delta}p^\rho p_h^\sigma (q^\gamma -k^\gamma)\eta^ \delta}{\tilde p_t\sqrt{\tilde \ly}}
\nn
&=&\frac{p_t\sqrt{\ly}\eta_2-2\varepsilon_{\rho\sigma\gamma\delta}p^\rho p_h^\sigma k^\gamma\eta^ \delta}{\tilde p_t\sqrt{\tilde \ly}},
\nn
\tilde \eta_3&=&-\frac{2M (q-k)\eta}{\sqrt{\tilde \ly}}
=\frac{\sqrt{\ly}\eta_3+2M k\eta}{\sqrt{\tilde \ly}},
\nn
\tilde \eta_t&=&\sqrt{-\tilde g^t_{\mu\nu}\eta^\mu\eta^\nu}=\sqrt{1-\tilde \eta_3^2}.
\label{tlet}
\ea
Here the tensors $\varepsilon^t_{\mu\nu}$ and $g^t_{\mu\nu}$ are modified as:
\ba
\tilde \varepsilon^t_{\mu\nu}&=&\frac{2\varepsilon_{\mu\nu\rho\sigma}p^\rho (q^\sigma-k^\sigma)}{\sqrt{(S_x-R)^2+4M^2(Q^2+\tau R)}},
\nn[2mm]
\tilde g^t_{\mu\nu}&=&\tilde \varepsilon^t_{\mu\rho}\tilde \varepsilon^{t\rho}_\nu=
\tilde g^\bot_{\mu\nu}
-\frac{4(Q^2+\tau R)\tilde p^\bot_\mu \tilde p^\bot_\nu}{(S_x-R)^2+4M^2(Q^2+\tau R)}.
\nn
\ea
For the integration of the expression (\ref{srfin}) over the real photon variables we need to express the terms containing $k$ in terms of
scalar quantities:
\ba
&\displaystyle
k\eta=\frac{R}{
2M\sqrt{\ly}}\biggl[(2M^2\tau-S_x)\eta_3
\nn
&\displaystyle
-
2M^2\eta_t\sqrt{(\tau_{max}-\tau)(\tau-\tau_{min})}\cos(\phi_\eta-\phi_k)\biggr],
\nn
&\displaystyle
\varepsilon_{\rho\sigma\gamma\delta}p^\rho p_h^\sigma k^\gamma\eta^ \delta=
\frac{R}{2\sqrt{\ly}}\biggl[(S_x-2\tau M^2)\eta_2
\nn&\displaystyle
+2M^2(\eta_3p_t\sin(\phi_h-\phi_k)-\eta_tp_l\sin(\phi_\eta-\phi_k))
\nn&\displaystyle
\times \sqrt{(\tau_{max}-\tau)(\tau-\tau_{min})}\biggr].
\ea

In the leading logarithmic approximation, the expressions (\ref{tlet})  split into two parts
corresponding to collinear radiation along the initial or final electron (historically
known as $s$- and $p$-peaks). Both contributions come from the region where $\phi_k$ is equal to zero, and values of $R$ and $\tau$ specific for $s$- and $p$-peak: i) $\tau \to\tau_s\equiv -Q^2/S$ and $R=(1-z_1)S$ for s-peak and ii) $\tau \to \tau_p\equiv Q^2/X$ and  $R=(z_2^{-1}-1)X$ for p-peak. Here dimensionless variables $z_1$ and $z_2$ reflect the remaining degree of freedom, i.e. photon energy, as follows $k \to k_{s,p}$ where
\ba
k_s=(1-z_1)k_1,\qquad
k_s=(z_2^{-1}-1)k_2.
\ea

Thus
\ba
\eta^{s,p}_1&=& \eta_t^s \cos( \phi_h^{s,p}- \phi_\eta^{s,p})=-\frac{p_h^\mu \eta^\nu  g^{s,p}_{\mu \nu}}{ p_{t\;s,p}}
\nn
&=&\frac 1{p_{t\;s,p}}\biggl[\eta_3 p_l+\eta_1 p_t
\nn
&&
-p_{l\;s,p}(\cos\phi_\eta \sin\theta_{s,p}\eta_t-\cos\theta_{s,p}\eta_3)\biggr],
\nn
 \eta^{s,p}_2&=& \eta_t^{s,p} \sin( \phi_h^{s,p}- \phi_\eta^{s,p})=-\frac{p_h^\mu \eta^\nu  \varepsilon^{s,p}_{\mu \nu}}{  p_{ts}}
\nn
&=&\frac {1}{p_{t\;s,p}}\biggl[(\cos\theta_{s,p}\eta_2-\sin\theta_{s,p}\sin\phi_h\eta_3)p_t
\nn&&
+\sin\theta_{s,p}\sin\phi_\eta\eta_tp_l\biggr],
\nn
 \eta^{s,p}_3&=&-\frac{2M q_{s,p}\eta}{\sqrt{\lambda_{Ys}}}
=
\cos\theta_{s,p}\eta_3-\cos\phi_\eta\sin\theta_{s,p}\eta_t.
\label{etsp1}
\ea
Here $q_s=z_1k_1-k_2$, $q_p=k_1-z_2^{-1}k_2$, $\theta_{s,p}$ are the angles between three-momenta $\bf q$ and ${\bf q}_{s,p}$ in the target rest frame reference
system,
\vspace*{2mm}
\ba
 \varepsilon^{s,p}_{\mu\nu}&=&\frac{2\varepsilon_{\mu\nu\rho\sigma}p^\rho q_{s,p}^\sigma}{\sqrt{\lambda_{Y\; s,p}}},
\nn[2mm]
 g^{s,p}_{\mu\nu}&=& \varepsilon^{s,p}_{\mu\rho} \varepsilon^{s,p\;\rho}_\nu,
\ea
and the others quantities from (\ref{etsp1}) are presented as
\ba
p_{t\;s,p}&=&\sqrt{-p_h^\mu p_h^\nu g^{s,p}_{\mu\nu}}=\sqrt{\frac{z^2S_x^2}{4M^2}-p_{l\;s,p}^2-m_h^2},
\nn
p_{ls}&=&\frac{z S_x(z_1S-X)-2M^2(z_1V_1-V_2)}{2M\sqrt{\ly^s}},\;
\nn
p_{lp}&=&\frac{z S_x(S-z_2^{-1}X)-2M^2(V_1-z_2^{-1}V_2)}{2M\sqrt{\ly^p}},\;
\nn
\ly^s&=&(z_1S-X)^2+4z_1M^2Q^2,
\nn
\ly^p&=&(S-z_2^{-1}X)^2+4z_2^{-1}M^2Q^2,
\nn[2mm]
\cos\theta_s&=&\frac{{\bf q}\cdot{\bf q}_s}{|{\bf q}||{\bf q}_s|}=\frac{(z_1S-X)S_x+2(z_1+1)M^2Q^2}{\sqrt{\ly\ly^s}},
\nn
\sin\theta_s&=&\sqrt{1-\cos\theta_s^2}
\nn
&=&\frac{2(1-z_1)M\sqrt{Q^2(SX-M^2Q^2)}}{\sqrt{\ly\ly^s}},
\nn
\cos\theta_p&=&\frac{{\bf q}\cdot{\bf q}_p}{|{\bf q}||{\bf q}_p|}=\frac{(S-z_2^{-1}X)S_x+2(1+z_2^{-1})M^2Q^2}{\sqrt{\ly\ly^p}},
\nn
\sin\theta_p&=&\sqrt{1-\cos\theta_p^2}
\nn
&=&\frac{2(z_2^{-1}-1)M\sqrt{Q^2(SX-M^2Q^2)}}{\sqrt{\ly\ly^p}}.
\ea
Here we take into account that
\ba
&\displaystyle
\sqrt{(\tau_{max}-\tau_s)(\tau_s-\tau_{min})}=\frac{\sqrt{Q^2(SX-M^2Q^2)}}{MS},
\nn
&\displaystyle
\sqrt{(\tau_{max}-\tau_p)(\tau_p-\tau_{min})}=\frac{\sqrt{Q^2(SX-M^2Q^2)}}{MX},
\nn
&\displaystyle
|{\bf q}|=\frac{\sqrt{\ly}}{2M},\qquad
|{\bf q}_{s,p}|=\frac{\sqrt{\ly^{s,p}}}{2M}.
\ea

Notice that the presented decomposition can be expressed through Born components $\eta_{1-3}$ (\ref{etab}) as a result of the three rotations
\ba
\left( \matrix{\eta_1^{s,p}\cr\eta_2^{s,p}\cr\eta_3^{s,p}\cr}\right)
=A^{s,p}_{\phi_h}A^{s,p}_\theta A_{\phi_h}
\left( \matrix{\eta_1 \cr \eta_2 \cr \eta_3\cr}\right)
\ea
\vspace{2mm}

The first rotation around $\bf q$
\ba
A_{\phi_h}=\left( \matrix{\cos\phi_h & -\sin\phi_h & \;\;0  \cr
\sin\phi_h & \;\cos\phi_h & \;\;0\cr
0  & 0 & 1 \cr} \right)
\ea
turns the basis from the hadronic plane to leptonic planes, as depicted in Fig.~\ref{fgtr}.

The second transformation in the leptonic plane
\ba
A^{s,p}_\theta=
\left( \matrix{
\cos\theta_{s,p} &\;\;0 &\;\; \sin\theta_{s,p} \cr
0&1 & 0 \cr
-\sin\theta_{s,p}  & 0 & \cos\theta_{s,p} \cr}\right)
\ea
change the direction of the axis $\bf z$ from $\bf q$ to ${\bf q}_{s,p}$.

At last rotation the basis turns to the true hadronic plane  
\ba
A^{s,p}_{\phi_h}=
\left( \matrix{
\cos\phi_h^{s,p} & \sin\phi_h^{s,p} & \;\;0 \cr
-\sin\phi_h^{s,p} & \;\cos\phi_h^{s,p} & \;\;0 \cr
0  & 0 & 1 \cr}\right)
\ea
whose changes were induced by real photon emission.

Here
\ba
\cos\phi_h^{s,p}&=&-\frac{k_1^\mu p_h^\nu g^{s,p}_{\mu\nu}}{k_t^{s,p}p_{t\;s,p}}
=\frac{p_t\cos\theta_{s,p} \cos\phi_h+p_l\sin\theta_{s,p} }{p_{t\;s,p}},
\nn
\sin\phi_h^{s,p}&=&-\frac{k_1^\mu p_h^\nu \varepsilon^{s,p}_{\mu\nu}}{k_t^{s,p}p_{t\;s,p}}
=\frac{p_t\sin\phi_h}{p_t^s},
\nn
k_t^{s,p}&=&\sqrt{-k_1^\mu k_1^\nu g^{s,p}_{\mu\nu}}=\sqrt{\frac{Q^2(SX-M^2Q^2)}{\ly^{s,p}}}.
\ea

As a result we obtained that
\ba
\eta^{s,p}_i=\sum\limits_{j=1}^3A^{s,p}_{ij}\eta_j,
\label{etsp2}
\ea
where  $A^{s,p}=A^{s,p}_{\phi_h}A^{s,p}_\theta A_{\phi_h}$ and
\ba
A^{s,p}_{11}&=&\frac{1}{p_{t\;s,p}}\biggl[(1-\cos^2\phi_h\sin^2\theta_{s,p})p_t
\nn&&\qquad
+\sin\theta_{s,p}\cos\theta_{s,p}\cos\phi_h p_l\biggr],
\nn
A^{s,p}_{12}&=&\frac{1}{p_{t\;s,p}}\biggl[\cos\phi_h\sin\phi_h\sin^2\theta_{s,p}p_t
\nn&&\qquad
-\sin\phi_h\cos\theta_{s,p}\sin\theta_{s,p}p_l\biggr],
\nn
A^{s,p}_{13}&=&\frac{\cos\phi_h\cos\theta_{s,p}\sin\theta_{s,p}p_t
+\sin^2\theta_{s,p}p_l}{p_{t\;s,p}},
\nn
A^{s,p}_{21}&=&
\frac{\sin\phi_h\sin\theta_{s,p} p_l}{p_{t\;s,p}},
\nn
A^{s,p}_{22}&=&
\frac{\cos\theta_{s,p} p_l+\cos\phi_h\sin\theta_{s,p} p_t}{p_{t\;s,p}},
\nn
A^{s,p}_{23}&=&
\frac{\sin\phi_h\sin\theta_{s,p} p_t}{p_{t\;s,p}},
\nn
A^{s,p}_{31}&=&\cos\phi_h\sin\theta^{s,p},
\nn
A^{s,p}_{32}&=&-\sin\phi_h\sin\theta^{s,p},
\nn
A^{s,p}_{33}&=&\cos\theta^{s,p}.
\ea

It is easy to show by direct analytical comparison that Eqs.~(\ref{etsp1}) and (\ref{etsp2}) are identical.

\section{Improved expression for $\theta_{53}^0$}
\label{th5}
The quantities $\theta_{ij}$ incoming into Eq.~(\ref{srfin}) are appeared as a result of  convolution of the leptonic radiative tensor presented by Eq.~ (25) of \cite{AI2019} with the hadronic structures $\tilde w^i_{\mu\nu}$. Generally, such a convolution includes the square of the leptonic propagators that is proportional to the square of the lepton mass. However, as presented in Appendix B of \cite{AI2019}, only the quantity
$\hat\theta^0_{53}$ does not obey this rule.
This situation occurs because of a specific form of the polarized part of the leptonic tensor we used to obtain Eq.~(26) of \cite{AI2019}:
\ba
L_{pR}^{\mu \nu }&=&4i\lambda_e\Biggl[
\Biggl\{
\Biggl(
\frac{2m^2F_{2+}-2(Q^2+2m^2)F_d}{R^2}
-\frac {\tau F_d}R \Biggr)
\nonumber\\&&\times
k_{1\alpha} k_{2\beta}
+
\Biggl(
\frac {\tau (F_d+F_{21})}R
+
\frac{1}{R^2}((Q^2+4m^2)F_d
\nonumber\\&&
-2m^2F_{22}-(Q^2+4m^2)F_{21})
\Biggr)
k_\alpha k_{1\beta}
\Biggr\}
\varepsilon^{\mu \nu \alpha \beta}
\nonumber\\&&
+2\frac{k_2^\mu F_d-(k^\mu+k_1^\mu)F_{21}}{R^2}
\varepsilon^{\nu \alpha \beta\gamma}k_{\alpha} k_{1\beta}k_{2\gamma}
\nonumber\\&&
-2\frac{k_2^\nu F_d-(k^\nu+k_1^\nu)F_{21}}{R^2}
\varepsilon^{\mu \alpha \beta\gamma}k_{\alpha} k_{1\beta}k_{2\gamma}
\Biggr].
\ea
After applying an identity
\ba
g_{\mu\nu}\varepsilon^{\alpha\beta\gamma\delta}&=&
g_{\mu\alpha}\varepsilon^{\nu\beta\gamma\delta}
+g_{\mu\beta}\varepsilon^{\alpha\nu\gamma\delta}
+g_{\mu\gamma}\varepsilon^{\alpha\beta\nu\delta}
\nn&&
+g_{\mu\delta}\varepsilon^{\alpha\beta\gamma\nu}
\ea
we can find that
\ba
L_{pR}^{\mu \nu \ast}&=&4i\lambda_e
\varepsilon^{\mu \nu \alpha \beta}
\Biggl[
\Biggl(
\frac{2m^2F_{2+}-2(Q^2+2m^2)F_d}{R^2}
\nonumber\\&&
-\frac {\tau F_d}R \Biggr)
k_{1\alpha} k_{2\beta}
+
\Biggl(\frac{(Q^2+4m^2)F_d-2m^2F_{22}}{R^2}
\nonumber\\&&
+\frac {\tau F_d+F_{1+}}{2R} \Biggr)k_\alpha k_{1\beta}
+
\Biggl(
\frac {\tau F_d-F_{1+}}{2R}
\nonumber\\&&
+
\frac{(Q^2+4m^2)F_d-2m^2F_{21}}{R^2}
\Biggr)k_\alpha k_{2\beta}
\Biggr].
\label{lrpim}
\ea
Therefore after convolution of the improved leptonic tensor (\ref{lrpim}) with $\tilde w^5_{\mu\nu}$
the quantity $\theta^0_{53}$ has an equivalent form
\ba
\theta^0_{53}&=&\frac{\lambda_e S}{\lambda_1\lambda_S}[\epa(\tau ((2\tau M^2-S_x)(Q^2+4m^2)
\nonumber\\&&
+2\tau (M^2Q^2-S X))F_d
-S_pQ^2F_{1+}+2\lambda_q)
\nonumber\\&&
+\frac{\varepsilon_\bot k}R(\tau (S_p(XV_1-SV_2)
\nonumber\\&&
+(z S_x^2-4M^2V_-)(Q^2+4m^2))F_d
\nonumber\\&&
+(S_pS_x(zQ^2+V_-)
-\lambda_q V_+)F_{1+})],
\label{th53}
\ea
that does not contain the square of the leptonic propagators $F_{22}$ and $F_{21}$.

\section{Exclusive structure functions.}
\label{exsf}
The most compact expression for the exclusive structure functions $H^{ex}_{ab}$
can be obtained through the so-called Chew-Goldberger-Low-Nambu amplitudes
\cite{CGLN} ${\cal F}_i$ in the following way:
\ba
f_1&=&\frac {4\sqrt{2\pi}W^2}{(W-M)\sqrt{\alpha r_1r_3}}{\cal F}_1
\nn
&=&\frac 1{2\sqrt{2\pi\alpha}}
\biggl[
A_1
+(W-M)A_4
\nn&&
+\frac{Q^2A_6+V_-(A_3-A_4)}{W-M}
\biggr],
\nn
f_2&=&\frac {4\sqrt{2\pi}W^2}{(W+M)\sqrt{\alpha r_2r_4}}{\cal F}_2
\nn
&=&\frac 1{2\sqrt{2\pi\alpha}}
\biggl[
-A_1+(W+M)A_4
\nn&&
+\frac{Q^2A_6+V_-(A_3-A_4)}{W+M}
\biggr],
\nn
f_3&=&\frac {8\sqrt{2\pi}W^3}{(W+M)r_3\sqrt{\alpha r_2r_4}}{\cal F}_3
\nn
&=&\frac 1{2\sqrt{2\pi\alpha}}
\biggl[
A_3-A_4+(W-M)A_2
\nn&&
+\frac{Q^2(A_2-2A_5)}{2(W+M)}
\biggr],
\nn
f_4&=&\frac {8\sqrt{2\pi r_3}W^3}{(W-M)((W^2+m_h^2-m_u^2)^2-4W^2m_h^2)\sqrt{\alpha r_1}}{\cal F}_4
\nn
&=&\frac 1{2\sqrt{2\pi\alpha}}
\biggl[
A_3-A_4-(W+M)A_2
\nn&&
-\frac{Q^2(A_2-2A_5)}{2(W-M)}
\biggr]
\nn
f_5&=&\frac {4\sqrt{2\pi r_1}W^2}{(W^2-M^2-Q^2)\sqrt{\alpha r_3}}{\cal F}_5
\nn
&=&\frac 1{8W\sqrt{2\pi\alpha}}
\biggl[
(W^2+m_h^2-m_u^2)(Q^2(A_2-2A_5)
\nn&&
+2(M+W)((W-M)A_2+A_3-A_4))
\nn&&
+V_-((M^2+Q^2-W^2)(2A_5-3A_2)
\nn&&
+4W(A_4-2WA_2-A_3)) -((M^2+Q^2-W^2)^2
\nn&&
+4Q^2W^2)A_2
+2r_1((W-M)(A_4-A_6)+A_1)
\biggr],
\nn
f_6&=&\frac {\sqrt{2\pi r_2}W^2}{(W^2-M^2-Q^2)\sqrt{\alpha r_4}}{\cal F}_6
\nn
&=&\frac 1{8W\sqrt{2\pi\alpha}}
\biggl[
\ly A_2-2r_2(A_1+(W+M)A_6)
\nn&&
+(W^2+m_h^2-m_u^2)(2Q^2A_5+2(W-M)A_3
\nn&&
+2MA_4+(2M^2-2W^2-Q^2)A_2)
\nn&&
+2(M(M^2+Q^2-W^2)-W(M^2-m_u^2+t))A_4
\nn&&
+V_-(2(W^2-M^2-Q^2)A_5
-4WA_3
\nn&&
+(3M^2+5W^2+3Q^2)A_2)
\biggr],
\ea
where
\ba
&\displaystyle
r_1=(W+M)^2+Q^2,\;
r_2=(W-M)^2+Q^2,\;
\nn
&\displaystyle
r_3=(W+m_u)^2-m_h^2,\;
r_4=(W-m_u)^2-m_h^2,
\ea

Taking into account that each $f_a$ has real and imaginary parts, i.e. representation $f_a=f_a^r+if_a^i$ where $f_a^{r,i} \in R$
we can construct the following combination  
\ba
f_{aa}=(f_a^r)^2+(f_a^i)^2,
\nn
f_{ab}^r=f_a^rf_b^r+f_a^if_b^i
\nn
f_{ab}^i=f_a^if_b^r-f_a^rf_b^i
\ea
for $a,b=1-6$ and $a<b$.

As a result the exclusive structure functions in orthogonal basis read:
\ba
&\displaystyle
H^{ex}_{00}+H^{ex}_{22}=\frac 1{W^2}\biggl[
2Q^2W^2\biggl(\frac {r_3} {r_1}f_{55}+\frac {r_4} {r_2}f_{66}\biggr )\qquad
\nn
&\displaystyle
\qquad
+\frac {r_5}{r_1r_2}((W^2-M^2)\ly f_{12}^r-4Q^2W^2f_{56}^r)
\nn
&\displaystyle
+\frac 12((W-M)^2r_1r_3f_{11}+(W+M)^2r_2r_4f_{22}) \biggr]
\nn
&\displaystyle
+\frac{2\eta_2 p_t}{W\sqrt{\ly}}[(W^2-M^2)\ly f_{12}^i+4Q^2W^2f_{56}^i],
\nn
&\displaystyle
\frac{H^{ex}_{11}-H^{ex}_{22}}{p_t^2}=\frac 1{2W^2}\biggl[
(W+M)^2r_2(4Wf_{23}^r+r_3f_{33})
\nn
&\displaystyle
+
2(M^2-W^2)r_5f_{34}^r
+
(W-M)^2r_1(4 W f_{14}^r+r_4f_{44})
\biggr]
\nn
&\displaystyle
+\frac{\eta_2}{W^2p_t\sqrt{\ly}}\biggl[
r_5((W-M)^2r_1f_{14}^i-(M+W)^2r_2f_{23}^i)
\nn
&\displaystyle
+
(W^2-M^2)\ly(
r_4f_{24}^i+2W(p_t^2f_{34}^i
-2f_{12}^i)
\nn&\displaystyle
-r_3f_{13}^i
)
\biggr],
\nn
&\displaystyle
H^{ex}_{22} =\frac 1{2W^2}\biggl[(W-M)^2r_1r_3f_{11}
+(W+M)^2r_2r_4f_{22}
\nn
&\displaystyle
+2(W^2-M^2)r_5f_{12}^r\biggr]
+\frac{2\eta_2p_t\sqrt{\ly}}W(W^2-M^2)f_{12}^i,
\nn
&\displaystyle
H_{01}^{ex\;r} =\frac {Qp_t}{W\sqrt{\ly}}\biggl[(W-M)(r_1(2Wf_{16}^r+r_4f_{46}^r)
\nn
&\displaystyle
-r_5f_{45}^r)+(W+M)(r_2(2Wf_{25}^r+r_3f_{35}^r)-r_5f_{36}^r)
\biggr]
\nn
&\displaystyle
+\frac{\eta_2Q}{r_1r_2W}\biggl[(W-M)(r_1r_5f_{16}^i-\ly(r_3f_{15}^i+2Wp_t^2f_{45}^i))
\nn
&\displaystyle
+(M+W)(\ly(r_4f_{26}^i+2Wp_t^2f_{36}^i)-r_2r_5f_{25}^i)
\biggr],
\nn
&\displaystyle
H_{01}^{ex\;i} =\frac {Qp_t}{W\sqrt{\ly}}\biggl[(W-M)(
r_5f_{45}^i-
r_1(2Wf_{16}^i
\nn
&\displaystyle
+r_4f_{46}^i)
)+(W+M)(r_5f_{36}^i-r_2(2Wf_{25}^i+r_3f_{35}^i))
\biggr]
\nn
&\displaystyle
+\frac{\eta_2Q}{r_1r_2W}\biggl[(W-M)(r_1r_5f_{16}^r-\ly(r_3f_{15}^r+2Wp_t^2f_{45}^r))
\nn
&\displaystyle
+(M+W)(\ly(r_4f_{26}^r+2Wp_t^2f_{36}^r)-r_2r_5f_{25}^r)
\biggr],
\nn
&\displaystyle
H_{02}^{ex\;r}=\frac {Q\eta_1}{r_1r_2W}\biggl[
(W-M)(r_3\ly f_{15}^i-r_1r_5f_{16}^i)
\nn
&\displaystyle
+(W+M)r_2(r_5 f_{25}^i-r_1r_4f_{26}^i)\biggr]
\nn
&\displaystyle
-\frac{2\eta_3Qp_t}{\sqrt{\ly}}\biggl[
(W+M)r_2f_{25}^i
+(W-M)r_1f_{16}^i\biggr],
\nn
&\displaystyle
H_{02}^{ex i}=\frac {Q\eta_1}{r_1r_2W}\biggl[
(W-M)(r_3\ly f_{15}^r-r_1r_5f_{16}^r)
\nn
&\displaystyle
+(W+M)r_2(r_5 f_{25}^r-r_1r_4f_{26}^r)\biggr]
\nn
&\displaystyle
-\frac{2\eta_3Qp_t}{\sqrt{\ly}}\biggl[
(W+M)r_2f_{25}^r
+(W-M)r_1f_{16}^r\biggr],
\nn
&\displaystyle
H_{12}^{ex\; r}=\frac {p_t\eta_1}{2W^2\sqrt{\ly}}\biggl[
(W^2-M^2)(r_1r_2(4Wf_{12}^i-r_4f_{24}^i)
\nn
&\displaystyle
+r_3\ly f_{13}^i)
+r_5((W+M)^2r_2f_{23}^i-(W-M)^2r_1f_{14}^i)\biggr]
\nn
&\displaystyle
-\frac{\eta_3p_t^2}{W}\biggl[
(W-M)^2r_1f_{14}^i
+(W+M)^2r_2f_{23}^i\biggr],
\nn
&\displaystyle
H_{12}^{ex\;i}=\frac {p_t\eta_1}{2W^2\sqrt{\ly}}\biggl[
(W^2-M^2)(
r_3\ly f_{13}^r-
r_1r_2r_4f_{24}^r)
\nn
&\displaystyle
+r_5((M+W)^2r_2f_{23}^r-(M-W)^2r_1f_{14}^r)\biggr]
\nn
&\displaystyle
-\frac{\eta_3}{2W^2r_1r_2}\biggl[
(W-M)^2r_1(2W\ly p_t^2f_{14}^r+r_1r_2r_3 f_{11})
\nn
&\displaystyle
+
(W+M)^2r_2(2W\ly p_t^2f_{23}^r+r_1r_2r_4 f_{22})
\nn
&\displaystyle
+
2 (W^2-M^2)r_5\ly f_{12}^r
\biggr],
\ea
with
\ba
&\displaystyle
r_5=W^2(M^2+m_h^2+m_u^2-Q^2-2t)
\nn
&\displaystyle
+(M^2+Q^2)(m_h^2-m_u^2)-W^4.
\ea

\end{document}